\documentclass[prd,showpacs,amsmath,amssymb,superscriptaddress,nofootinbib,showkeys]{revtex4}
\usepackage{amssymb}
\usepackage{amsmath}
\usepackage[cp1251]{inputenc}
\usepackage[T2A]{fontenc}
\usepackage{graphicx}

\begin{document}
\title{The magnetic universe in the theory of gravity with
non-minimal kinetic coupling of a scalar field}

\author{Ruslan K. Muharlyamov}
\email{rmukhar@mail.ru} \affiliation{Department of General
Relativity and Gravitation, Institute of Physics, Kazan Federal
University, Kremlevskaya str. 18, Kazan 420008, Russia}

\author{Tatiana N. Pankratyeva}
\email{ghjkl.15@list.ru} \affiliation{Department of Higher
Mathematics, Kazan State Power Engineering University,
Krasnoselskaya str. 51, Kazan 420066, Russia}

\begin{abstract}
In this paper, we obtain cosmological solutions within the framework of Bianchi I model with a unidirectional magnetic field in the theory of gravity with non-minimum kinetic coupling of a scalar field with curvature ($\eta\cdot(\phi/2)\cdot G_{\mu\nu}\,\nabla^\mu \nabla^\nu
\phi$). We believe that the solutions are applicable to the period before and during primary inflation. In the previous work [R. K. Muharlyamov, T. N. Pankratyeva, S. O. A. Bashir,  JCAP05(2025)018] we considered such a problem for a special case -- the locally rotationally symmetric Bianchi I space-time ($a_1=a_2$). Here we generalize the results for different scale factors $a_1\neq a_2 \neq a_3$. There are distinctive properties of the models and a number of interesting properties are preserved.

The main conclusion remains valid. For $l=1+\varepsilon\eta \Lambda/\mu>0$, the model has the necessary properties: isotropization during expansion, rapid transition to inflationary expansion, absence of ghost and Laplace instabilities. Here, $\Lambda$ is the cosmological constant,  $\mu=M^2_{PL}$ is the Planck mass squared, the sign $\varepsilon=\pm 1$ defines the canonical scalar field and the phantom field, respectively.

The differences in the model appear in the neighborhood of the starting point and at times of several units $h_\eta\cdot t$ from it ($h_\eta=\sqrt{\frac{\varepsilon}{3\eta}}$). In the previous model with $l>0$, the Universe is described by the Kasner metric in the neighborhood of the initial point. In the generalized model, there is no the Kasner metric. In the presented model, the anisotropy level is a non-monotonic function. There is a rebound point on the plot. This is the non-standard  behavior.
But, in the end, the process of isotropization occurs.

\end{abstract}



\maketitle

\section{Introduction}

The observational cosmology has allowed us to deepen our knowledge about the Universe \cite{Komatsu,Susuki,Hinshaw,PAR,DESI1,DESI2, BICEP, Atacama0, Atacama1,Aghanim}.
An important discovery was the accelerated expansion of the Universe and evidence of the existence of  dark matter. The theoretical description of these facts is one of the main motives for modifications of the theory of gravity. The most successful extension of the theory of gravity without global complications is the Horndeski gravity (HG) \cite{Horndeski}. The field equations within the framework of this theory have derivative order no higher than second. In this paper we restrict ourselves to the subclass of HG with a non-minimal kinetic coupling (NMKC) of a scalar field $\phi$ with the Einstein tensor with the action density
\begin{eqnarray}\label{lagr1} L_H=\sqrt{-g}\left(\frac{\mu R}{2}+G_2(X,\phi)+\frac{\eta\phi}{2}G_{\mu\nu}\,\nabla^\mu \nabla^\nu
\phi\right) \,,\end{eqnarray}
where $g$ is the determinant of metric tensor
$g_{\mu\nu}$; $R$ is the Ricci scalar and $G_{\mu\nu}$ is the
Einstein tensor;  $\mu=M^2_{PL}$ is the Planck mass squared. We will assume $G_2(X,\phi)=\varepsilon X-\Lambda$; $\Lambda$ is the cosmological constant  with the dimension $[\Lambda]=[L^{-4}]$; the sign $\varepsilon=\pm 1$ defines the canonical scalar field and the phantom field, respectively; $X=-\frac{1}{2}\nabla^\mu\phi \nabla_\mu\phi$ is the kinetic
term. NMKC involves the additional dimensional parameter $\eta$ with dimension of {\it (length)$^2$}, $[\eta]=[L^2]=[T^2]=[M^{-2}]$. We assume $c=\hbar=1$. Here we obtain cosmological solutions with the scalar field and the magnetic field, that can describe the early times of the Universe evolution. For this reason we do not include ordinary matter. The electromagnetic contribution to the action density is
\begin{equation}\label{lagrF}L_{F}=-\frac{\sqrt{-g}}{4}F_{\mu\nu}F^{\mu\nu},\end{equation}
where $F_{\mu\nu}$ is the electromagnetic field.

 A natural modification of General Relativity are models that take into account the non-minimal connection of the scalar field with curvature. For example, the Lagrangian of a gravity theory with a non-minimally coupled scalar field $\phi$ has the general form $F(\phi,R)$.  Although the freedom to choose the function $F(\phi,R)$ leads to an infinite number of different scalar-tensor theories, it is known that, using suitable conformal transformations, any such theory can be transformed into a theory of a minimally coupled scalar field with some effective potential \cite{Maeda}.  Another approach to constructing modified scalar-tensor theories of gravity is represented by models that take into account the non-minimal connection between the derivatives of the scalar field with the curvature of spacetime.  In 1993, Amendola \cite{Amendola} showed that the scalar-tensor theory with non-minimal kinetic coupling is not conformally equivalent to Einstein's theory with a minimally coupled scalar field with potential, and hence its consequences may have qualitatively new features.  NMKC may appear in some Kaluza-Klein theories \cite{Shafi1,
Shafi2}.   NMKC offers a variety of effects for astrophysical objects such as wormholes \cite{Sushkov,Korolev}, black holes \cite{Rinaldi,Minamitsuji,Kobayashibb,Babichev}, neutron stars \cite{Cisterna,Maselli,Silva,Kashargin}. NMKC is also interesting from a cosmological point of view \cite{Sush2009,Sar2010,Sush2012,Sush2020,SkuSusTop2013,MatSus2015,StaSusVol2016,StaSusVol2019}.
Firstly, NMKC offers a different inflation mechanism without potential. Secondly, there is a screening property.  At early times of the Universe evolution, the $\Lambda$-term and matter are screened. This effect may disappear over time. These properties make it possible to describe the change of cosmological epochs with one model. There are interesting anisotropic effects. For example, in work \cite{StaSusVol2019},  it was found that anisotropy is screened at early time within the Bianchi I (BI) space-time model. This gives a non-monotonic behavior of the anisotropy level.

The existence of a large-scale magnetic field in the Universe has prompted many researchers to study models of the Universe with the magnetic field \cite{Doroshkevich,Thorne,Jacobs,Salimyx,Horwood,Bronnikov,Watanabe,Soda,Do,Nguyen,Petriakova,Muharlyamov24,Muharlyamov241} and  the Yang-Mills field \cite{Moniz00, Moniz0,Dyadichev0,Galtsov,Moniz1}.  In this paper we adhere to the hypothesis of the primary origin of the magnetic field. According to the hypothesis, the field arises at the very initial stages of the evolution of the Universe.  The model of cosmic magnetogenesis is presented, for example, in paper \cite{Moniz1} (additional references can be found there).  The scalar field can explain the phenomenon of cosmic acceleration. Unlike the
electromagnetic field, direct observation of a scalar field is not available (only consequences). From this position, their joint consideration in physical models significantly expands the possibilities of understanding and interpreting observational data associated with the dark sector of the Universe. It is natural to assume that the magnetic field has a certain influence on the phenomenon of cosmic acceleration.  An example of the pure influence of vector fields on the dynamics of cosmological models can be found in the works \cite{Moniz0, Dyadichev0}.

The interaction of physical fields will be studied within the framework of the popular model BI \cite{Muharlyamov0,SushkovStar,Muharlyamov2,Muharlyamov3,Koussour,Koussour1,Akarsu,Sarmah,Hawking0,Hu,Momeni,Momeni1,Mehran,Marco}. Planck’s data on the temperature and the polarization of the cosmic microwave background (CMB) radiation indicate that the Universe is "imperfectly"\, isotropic \cite{Saadeh}. The presence of anomalies  on large scales of the CMB radiation can be explained by the existence of an anisotropic early Bianchi Universe \cite{Komatsu,Ellis, Thorsrud}.

In the previous work \cite{Muhryamov4q}, we have studied in detail the cosmological solutions in theory (1), (2) for the special case --  the locally rotationally symmetric BI space-time ($a_1=a_2$).  Here we generalize the results for different scale factors $a_1\neq a_2 \neq a_3$. We consider the issue of isotropization of the model, primary inflation, the behavior of the model near the singularity and the features of the evolution of the magnetic field. We will look for models of only an expanding Universe. We limit ourselves to the period before and during primary inflation.

\section{Field equations}\label{sec2}

We will consider the homogeneous and anisotropic space-time BI:
\begin{equation}
 ds^2 = -dt^2 + a^2_1(t)dx_1^2 + a^2_2(t)dx_2^2 + a^2_3(t)dx_3^2. \label{met0}
\end{equation}

In this metric, the gravity equations have the form
\begin{eqnarray} \label{00}
 G^0_0\left(\mu+\frac{3\eta \dot{\phi}^2}{2}\right) = G_2 -
G_{2X}\dot{\phi}^2+T^{({\rm em})0}_{0}\,,\end{eqnarray}
\begin{eqnarray} \label{ii}G^{i}_{i}\left(\mu+\frac{\eta
\dot{\phi}^2}{2}\right)-(H_{j}+H_{k})\frac{d}{dt}\left(\mu+\frac{\eta
\dot{\phi}^2}{2}\right) =G_2 +T^{({\rm
em})i}_{i}  \,.
\end{eqnarray}
Here the dot denotes  the $t$-derivative, one has $H_i=\dot {
a}_i/{ a}_i$, and the average Hubble parameter is
$H=\dfrac{1}{3}\sum\limits_{i=1}^3 H_i\equiv \dot{ a}/{ a}$ with
${ a}=({ a}_1{ a}_2{ a}_3)^{1/3}$ -- the geometric mean  scale factor; $T^{({\rm em})\mu}_{\nu}$ -- the stress–energy tensor of the electromagnetic field.  The Einstein tensor components
are
\begin{eqnarray}
&&G^0_0=-\left(H_1H_2 +H_2H_3 +H_3H_1\right)\,, \\
&&G^{i}_{i}=-\left(\dot{H}_{j} +\dot{H}_{k} +H_{j}^2 +H_{k}^2
+H_{j}H_{k}\right)\,,
\end{eqnarray}
where the triples of indices $\{i,j,k\}$ take values $\{1,2,3\}$,
$\{2,3,1\}$, or $\{3,1,2\}$.

Varying the action (\ref{lagr1}) with respect to $\phi(t)$ yields the equation
\begin{eqnarray}\label{scalar}
\frac{1}{a^3}\frac{d}{dt}\left({ a^3}\dot{\phi}\, \Big[ G_{2X}+\eta G^0_0
\Big]\right)=G_{2\phi}.
\end{eqnarray}
Electromagnetic equations have the form $\partial_\mu[a^3F^{\mu\nu}]=0$ and there is the Bianchi identity $\nabla_\mu F_{\nu\alpha}+\nabla_\alpha F_ {\mu\nu}+\nabla_\nu F_{\alpha\mu}=0$.
We assume that there is the magnetic field
having the same direction  $x_3$.
The magnetic field corresponds to non-zero components $F_{\gamma\delta}$: $
F_{21}=-F_{12}=q_m$, where $q_m$ is constant.
The magnetic field strength is determined by the equality
\begin{eqnarray}\label{mag stren}
B^2=F_{21}F^{21}=\frac{q^2_m}{a^2_1a^2_2}\,.
\end{eqnarray}

The stress–energy tensor of the electromagnetic field is given by the following expression
\begin{equation}
T^{({\rm em})\mu}_{\nu}= -\frac{1}{4}\delta^\mu_\nu
F_{\gamma\delta}F^{\gamma\delta} + F_{\nu \beta} F^{\mu
\beta} \,. \label{em1}
\end{equation}
The tensor $T^{({\rm em})\mu}_{\nu}$ has non-zero components:
\begin{eqnarray}\label{tem}T^{({\rm em})0}_{0}=T^{({\rm em})3}_{3}=-T^{({\rm em})1}_{1}=-T^{({\rm em})2}_{2}=-\mathcal{E}^{({\rm em})}\,,\end{eqnarray}
where $\mathcal{E}^{({\rm em})}$ -- the magnetic field energy density:
\begin{equation}\label{eem}\mathcal{E}^{({\rm em})}=\frac{1}{2}B^2=\frac{q_m^2}{2a^2_1a^2_2}.\end{equation}

Let's consider the following parametrization of three scalar factors:
\begin{equation}
ds^2 =
-dt^2+e^{2\alpha(t)}[e^{2(\beta_{+}+\sqrt{3}\beta_{-})}dx_1^2+e^{2(\beta_{+}-\sqrt{3}\beta_{-})}dx_2^2+e^{-4\beta_+}dx_3^2].
\label{met}
\end{equation}
Then the Hubble parameters are given by
\begin{equation}\label{H}
H_1=H+\dot{\beta}_{+}+\sqrt{3}\dot{\beta}_{-}\,,\,
H_2=H+\dot{\beta}_{+}-\sqrt{3}\dot{\beta}_{-}\,, \,
H_3=H-2\dot{\beta}_{+}\,.
\end{equation}
The functions $e^{\beta_{+}\pm\sqrt{3}\beta_{-}}$, $e^{-2\beta_+}$ represent  deviations from isotropy,
while $e^{\alpha}=a(t)$ is the isotropic part.

In the previous work \cite{Muhryamov4q}, we studied case $\beta_{-}=0$ ($a_1=a_2$) that corresponds the locally rotationally symmetric
(LRS) Bianchi-I space-time.

As function $G_2(X,\phi)$, we take
\begin{equation} \label{g2} G_2=\varepsilon X-\Lambda, \end{equation}
which corresponds to the standard inflation theory.

In view of (\ref{tem}), (\ref{eem}), (\ref{met}), (\ref{H}) and (\ref{g2}), the system of field equations (\ref{00}), (\ref{ii}), (\ref{scalar}) has the consequences
\begin{equation} \label{lr00}3\big(\dot{\alpha}^2-\sigma^2\big)\left(\mu+\frac{3\eta \dot{\phi}^2}{2}\right)
=\frac{q^2_m}{2}\cdot e^{-4(\alpha+\beta_+)}
+\varepsilon\frac{\dot{\phi}^2}{2}+\Lambda,\end{equation}
$$\left(\mu+\frac{\eta
\dot{\phi}^2}{2}\right)\big(2\ddot{\alpha}+3\dot{\alpha}^2 +
3\sigma^2\big)+2\dot{\alpha}\frac{d}{dt}\left(\mu+\frac{\eta
\dot{\phi}^2}{2}\right)=$$\begin{equation} \label{lrii}=-\frac{q^2_m}{6}\cdot e^{-4(\alpha+\beta_+)}-\varepsilon\frac{\dot{\phi}^2}{2}+\Lambda,\end{equation}
\begin{equation} \label{lrb}e^{-3\alpha}\cdot\frac{d}{dt}\left[\left(\mu+\frac{\eta
\dot{\phi}^2}{2}\right)e^{3\alpha}\dot{\beta_+}\right]=\frac{q^2_m}{3}\cdot e^{-4(\alpha+\beta_+)},\end{equation}
\begin{equation}\label{lrb1}\left(\mu+\frac{\eta
\dot{\phi}^2}{2}\right)\dot{\beta}_{-} =c_\beta\cdot e^{-3\alpha}
\end{equation}
\begin{eqnarray}\label{scalar1}
\dot{\phi}\, \Big[ \varepsilon
-3\eta \big(\dot{\alpha}^2-\sigma^2\big)\Big]=C_{\phi}e^{-3\alpha}\,,\end{eqnarray}
where $C_{\phi}$ is the scalar charge, $c_\beta$ --integration constant,  $\sigma^2=\dot{\beta_+}^2+\dot{\beta_-}^2$ -- the shear scalar.
The magnetic field energy density is
\begin{equation}\mathcal{E}^{({\rm em})}=\frac{q^2_m}{2}\cdot e^{-4(\alpha+\beta_+)}.\end{equation}
The system (\ref{lr00})-(\ref{scalar1}) contains four independent equations.

Further, we put $C_{\phi}=0$.  In case $\dot{\phi}=0$, we obtain the system $$3\mu\big(\dot{\alpha}^2-\sigma^2\big)=\frac{q^2_m}{2}\cdot e^{-4(\alpha+\beta_+)}
+\Lambda,$$
\begin{equation}\mu e^{-3\alpha}\cdot\frac{d[e^{3\alpha}\dot{\beta_+}]}{dt}=\frac{q^2_m}{3}\cdot e^{-4(\alpha+\beta_+)},\,\, \mu\dot{\beta}_{-} =c_\beta\cdot e^{-3\alpha},\end{equation}
which corresponds to General Relativity. At late time, the Universe expansion is driven by $\Lambda$.

The second branch ($\dot{\phi}\neq0$) is determined by another consequence from (\ref{scalar1}):
\begin{equation}\label{dotbeta}\dot{\alpha}^2=\sigma^2+\frac{\varepsilon}{3\eta}.\end{equation}
Further we will consider only this branch.

Values $\dot{\beta_{\pm}}=q_m=0$ correspond to isotropic space-time and the Universe changes according to the law de Sitter or anti-de Sitter with parameters $\dot{\alpha}=\pm\sqrt{\frac{\varepsilon}{3\eta}}$ ($\frac{\varepsilon}{3\eta}>0$). The $\Lambda$-term is screened and does not contribute to the evolution of the Universe. In the anisotropic case ($\dot{\beta_{\pm}}\neq 0$) the question of screening remains open.
Nevertheless,  it can be said that in the presence of $\dot{\beta_{\pm}}\rightarrow 0$,  the geometric mean behavior of the Universe is determined by parameter $\sqrt{\frac{\varepsilon}{3\eta}}$:  $a(t)\propto e^{\pm\sqrt{\frac{\varepsilon}{3\eta}}\cdot t}$.

We will use the following criterion for isotropization of the cosmological model:
\begin{equation}\label{critizotrop} \frac{\sigma^2}{\dot{\alpha}^2}\rightarrow 0 \,\,\, \text{as} \,\,\, \alpha \rightarrow +\infty.\end{equation}
Ratio $\sigma_\alpha\equiv|\sigma/\dot{\alpha}|$ determines the  anisotropy level. It is found from the equation (\ref{dotbeta}):
\begin{equation}\label{dbdf}\sigma_\alpha^2=1-\frac{\varepsilon}{3\eta}\cdot\frac{1}{\dot{\alpha}^2}.\end{equation}

A necessary but not sufficient condition for the fulfillment of (\ref{critizotrop}) is the inequality
\begin{equation}\label{ettavar}\frac{\varepsilon}{\eta}>0.\end{equation}
The isotropization condition (\ref{critizotrop}) will be satisfied if $\dot{\alpha}^2$ is a decreasing function to the value $\frac{\varepsilon}{3\eta}$ at $t\rightarrow +\infty$.
Parameters $\varepsilon$ and $\eta$ have the same sign, so we can introduce the parameter
\begin{equation}\label{hett12} h_\eta=\sqrt{\frac{\varepsilon}{3\eta}}.\end{equation}
Therefore, equality (\ref{dbdf}) will take the form
\begin{equation}\label{alf}\sigma_\alpha^2=1-\frac{h_\eta^2}{\dot{\alpha}^2}\end{equation}
and there are restrictions:
\begin{equation}\label{bnaal}\sigma_\alpha<1,\end{equation}
\begin{equation}\label{h1}|H|=|\dot{\alpha}|\geq h_\eta,\end{equation}
that is, the anisotropy level $\sigma_\alpha$ is bounded and there is a lower non-zero limit to the Hubble parameter $|H|$.

Function $\dot{\phi}^2$ is found from equations (\ref{lr00}) and (\ref{dotbeta}):
$$\dot{\phi}^2=\varepsilon\mu\left[3(\varepsilon_\Lambda h_\Lambda^2-h_\eta^2)+\frac{\mathcal{E}^{({\rm em})}}{\mu}\right]=$$
\begin{equation}\label{phem}=\varepsilon\mu\left[3(\varepsilon_\Lambda h_\Lambda^2-h_\eta^2)+\frac{q^2_m}{2\mu}\cdot e^{-4(\alpha+\beta_+)}\right],\end{equation}
where $h_\Lambda=\sqrt{|\Lambda|/(3\mu)}$, $\varepsilon_\Lambda=\text{sign}(\Lambda)$.
Parameters $h_\eta$, $h_\Lambda$ have the dimension of the Hubble parameter: $[h_\eta]=[h_\Lambda]=[H]=[T^{-1}]=[L^{-1}]$, and
$[q_m]=[M^2]=[L^{-2}]$. From inequality (\ref{phem}) it follows that not all values of parameters
$\varepsilon$, $h_\eta$, $\varepsilon_\Lambda$, $h_\Lambda$ allow the density $\mathcal{E}^{({\rm em})}\rightarrow +\infty , \,\, 0$.

Substituting $\dot{\phi}^2$, $\sigma^2$  into equations (\ref{lrii}),  (\ref{lrb}), (\ref{lrb1}), we get
$$ e^{-3\alpha}\frac{d}{dt}\left\{\left[l+\frac{q^2_m}{6\mu h_\eta^2}\cdot e^{-4(\alpha+\beta_+)}\right]\dot{\alpha}e^{3\alpha}\right\}=$$
\begin{equation}=3h_\eta^2\left[l-\frac{q^2_m}{18\mu h_\eta^2}\cdot e^{-4(\alpha+\beta_+)}\right],\label{lriis}\end{equation}
\begin{equation} \label{lrbsj}e^{-3\alpha}\cdot\frac{d}{dt}\left\{\left[l+\frac{q^2_m}{6\mu h_\eta^2}\cdot e^{-4(\alpha+\beta_+)}\right]\dot{\beta_+}e^{3\alpha}\right\}=3h_\eta^2\cdot\frac{2q^2_m}{9\mu h_\eta^2}\cdot e^{-4(\alpha+\beta_+)},\end{equation}
\begin{equation}\label{lrb1aa}\dot{\beta}_{-} =\frac{2c_\beta}{\mu \cdot e^{3\alpha}\left(l+\frac{q^2_m}{6\mu h_\eta^2}\cdot e^{-4(\alpha+\beta_+)}\right)},\end{equation}
where
\begin{equation}\label{l} l\equiv 1+\frac{\varepsilon\eta \Lambda}{\mu}= 1+\varepsilon_\Lambda\cdot\left(\frac{h_\Lambda}{h_\eta}\right)^2.\end{equation}
The system (\ref{lriis}), (\ref{lrbsj}), (\ref{lrb1aa})  contains two independent equations. As independent equations we choose (\ref{lrb1aa}) and the combination (\ref{lriis})+(\ref{lrbsj}):
$$e^{-3\alpha}\frac{d}{dt}\left\{\left[l+\frac{q^2_m}{6\mu h_\eta^2}\cdot e^{-4(\alpha+\beta_+)}\right]e^{3\alpha}(\dot{\alpha}+\dot{\beta_+})\right\}=$$
\begin{equation}\label{lagt}=3h_\eta^2\left[l+\frac{q^2_m}{6\mu h_\eta^2}\cdot e^{-4(\alpha+\beta_+)}\right].\end{equation}
Taking into account equality (\ref{lrb1aa}), we rewrite equation (\ref{dotbeta}):
\begin{equation}\label{dkpo1}\dot{\alpha}^2=\dot{\beta_+}^2+\frac{4c_\beta^2}{\mu^2 \cdot e^{6\alpha}\left(l+\frac{q^2_m}{6\mu h_\eta^2}\cdot e^{-4(\alpha+\beta_+)}\right)^2}+h_\eta^2.\end{equation}
The system (\ref{lrb1aa}), (\ref{lagt}), (\ref{dkpo1}) defines functions $\alpha(t)$, $\beta_{\pm}(t)$.

Next, instead of function $\beta_+(t)$, we introduce a dimensionless function
\begin{equation}s(t)=\frac{q^2_m}{6|l|\mu h_\eta^2}\cdot e^{-4(\alpha+\beta_+)}>0,\end{equation}
which is associated with the magnetic field energy density $\mathcal{E}^{({\rm em})}$:
\begin{equation}s(t)=\frac{1}{3|l|\mu h_\eta^2}\cdot\mathcal{E}^{({\rm em})}.\end{equation}
We will replace the time coordinate $t$ with the dimensionless $\tau=h_\eta\cdot t$.
The system (\ref{lriis}), (\ref{lrbsj}), (\ref{lrb1aa}) will change as follows
\begin{equation}\label{mmsd1}(\beta_-)'_\tau=\frac{k}{|l|}\cdot\frac{1}{e^{3\alpha}\left(\varepsilon_l+s\right)},\,\, k\equiv\frac{2c_\beta}{\mu h_\eta},\end{equation}
\begin{equation}\label{mmsd2}\frac{d}{d\tau}\left\{\left(\varepsilon_l+s\right)e^{3\alpha}\cdot\frac{s'_\tau}{s}\right\}=-12 e^{3\alpha}\left(\varepsilon_l+s\right),\end{equation}
\begin{equation}\label{mmsd3}-8\alpha'_\tau\cdot\frac{s'_\tau}{s}=16+\left(\frac{s'_\tau}{s}\right)^2+\frac{16k^2}{|l|^2 \cdot e^{6\alpha}\left(\varepsilon_l+s\right)^2},\end{equation}
where $\varepsilon_l=\text{sign}(l)$. We will need a consequence of the equation (\ref{mmsd3}):
\begin{equation}\label{mmsd4}\frac{s'_\tau}{4s\alpha'_\tau}+1=\pm \sqrt{1-\frac{1}{\alpha'^2_\tau}-\frac{k^2}{|l|^2\cdot e^{6\alpha}\left(\varepsilon_l+s\right)^2}
\cdot\frac{1}{\alpha'^2_\tau}}.\end{equation}
In the previous work \cite{Muhryamov4q}, we studied case $k=0$ ($a_1=a_2$) that corresponds the
LRS Bianchi-I space-time.

The function $\dot{\phi}^2$ is expressed in terms of $s$:
\begin{equation}\label{phemlp}(\phi'_\tau)^2=3\varepsilon\mu\left(-1+\varepsilon_\Lambda \left(\frac{h_\Lambda}{h_\eta}\right)^2+\left|1+\varepsilon_\Lambda\cdot\left(\frac{h_\Lambda}{h_\eta}\right)^2\right|\cdot s\right).\end{equation}
From inequality $\dot{\phi}^2\geq 0$ follow restrictions on parameters $\varepsilon$, $h_\Lambda$, $h_\eta$, $\varepsilon_\Lambda$.

Square of the  anisotropy level $\sigma_\alpha$ takes the form
$$\sigma_\alpha^2=\frac{1}{\dot{\alpha}^2}\cdot(\dot{\beta_+}^2+\dot{\beta_-}^2)=$$\begin{equation}=\left(\frac{s'_\tau}{4s\alpha'_\tau}+1\right)^2+\frac{k^2}{|l|^2 \cdot e^{6\alpha}\left(\varepsilon_l+s\right)^2}
\cdot\frac{1}{\alpha'^2_\tau}.\label{levelsd4}\end{equation}
Here we used the equality
\begin{equation}\label{ajn}(\beta_+)'_\tau=-\alpha'_\tau-\frac{s'_\tau}{4s}.\end{equation}
We assumed that $l\neq 0$. Case $l=0$ will be considered separately.

\section{Case $l=0$}
Value $l=0$ corresponds to $\Lambda<0$, $h_\eta=h_\Lambda$. In this case, the system of equations takes the form:
\begin{equation}(\beta_-)'_\tau=\frac{k}{e^{3\alpha}s_1},\end{equation}
\begin{equation}\frac{d}{d\tau}\left\{e^{3\alpha}\cdot s'_{1\tau}\right\}=-12 e^{3\alpha}s_1,\end{equation}
\begin{equation}\label{mmsd3a}-8\alpha'_\tau\cdot\frac{s'_{1\tau}}{s}=16+\left(\frac{s'_{1\tau}}{s}\right)^2+\frac{16k^2}{e^{6\alpha}s_1^2},\end{equation}
where $s_1=\frac{1}{3\mu h_\eta^2}\cdot\mathcal{E}^{({\rm em})}=\frac{q^2_m}{6\mu h_\eta^2}\cdot e^{-4(\alpha+\beta_+)}$.
We will write the last two equations in a different form:
\begin{equation}\frac{d}{d\tau}\left\{(e^{3\alpha}s_1)'_{\tau}-3\alpha'_\tau(e^{3\alpha}s_1)\right\}=-12 (e^{3\alpha}s_1),\end{equation}
\begin{equation}15\alpha'_\tau(e^{3\alpha}s_1)=(e^{3\alpha}s_1)'_{\tau}\pm 4\sqrt{[(e^{3\alpha}s_1)'_{\tau}]^2+15[(e^{3\alpha}s_1)^2+k^2]}.\end{equation}
Eliminating $\alpha'_\tau(e^{3\alpha}s_1)$ from the system, we get
\begin{equation}(e^{3\alpha}s_1)''_{\tau\tau}+15(e^{3\alpha}s_1)=0\end{equation}
and therefore
$$e^{3\alpha}s_1=c_1\sin(\sqrt{15}\cdot\tau)+c_2\cos(\sqrt{15}\cdot\tau)=$$
\begin{equation}=\sqrt{c_1^2+c_2^2}\sin(\sqrt{15}\cdot\tau+\omega),\end{equation}
\begin{equation}e^{3\alpha}s_1\alpha'_\tau=(15)^{-1/2}\sqrt{c_1^2+c_2^2}\left(\cos(\sqrt{15}\cdot\tau+\omega)\pm 4\sqrt{1+\frac{k^2}{c_1^2+c_2^2}}\,\right).\end{equation}

where $\cos\omega=\frac{c_1}{\sqrt{c_1^2+c_2^2}}$, $\sin\omega=\frac{c_2}{\sqrt{c_1^2+c_2^2}}$.
Due to condition $e^{3\alpha}s_1\geq 0$, we restrict ourselves to the interval $\sqrt{15}\cdot\tau\in [-\omega,\pi-\omega]$ with period $2\pi$.
After simple calculations, we obtain the Hubble parameter $\alpha'_\tau$, the scale factor $a(\tau)$, $s_1(\tau)$ and $e^{\beta_-}$, respectively:
\begin{equation}\sqrt{15}\alpha'_\tau=\cot(\sqrt{15}\cdot\tau+\omega)\pm\frac{4\cdot\sqrt{1+\frac{k^2}{c_1^2+c_2^2}}}{\sin(\sqrt{15}\cdot\tau+\omega)},\end{equation}
\begin{equation}\label{lofact}a=e^{\alpha}=a_0\left\{\sin(\sqrt{15}\cdot\tau+\omega)\left[\tan(\sqrt{15}\cdot\tau/2+\omega/2)\right]^{\pm4\cdot\sqrt{1+\frac{k^2}{c_1^2+c_2^2}}}\right\}^{\frac{1}{15}},\end{equation}
\begin{equation}s_1=a_0^{-3}\sqrt{c_1^2+c_2^2}\left[\sin(\sqrt{15}\cdot\tau+\omega)\left[\tan(\sqrt{15}\cdot\tau/2+\omega/2)\right]^{\mp\sqrt{1+\frac{k^2}{c_1^2+c_2^2}}}\right]^{\frac{4}{5}},\end{equation}
\begin{equation}e^{\beta_-}=b_0\left[\tan(\sqrt{15}\cdot\tau/2+\omega/2)\right]^{\frac{k}{\sqrt{15}\cdot\sqrt{c_1^2+c_2^2}}}.\end{equation}

Scale factors $a_i$ have the form:
$$a_1=a_0^{3/4}b_0^{\sqrt{3}}(c_1^2+c_2^2)^{-1/8}\left(\frac{q^2_m}{6\mu h_\eta^2}\right)^{1/4}\left[\sin(\sqrt{15}\cdot\tau+\omega)\right]^{-1/5}\times$$$$
\times\left[\tan(\sqrt{15}\cdot\tau/2+\omega/2)\right]^{\pm\frac{1}{5}\cdot\sqrt{1+\frac{k^2}{c_1^2+c_2^2}}+\frac{k}{\sqrt{5}\sqrt{c_1^2+c_2^2}}},$$
$$a_2=a_0^{3/4}b_0^{-\sqrt{3}}(c_1^2+c_2^2)^{-1/8}\left(\frac{q^2_m}{6\mu h_\eta^2}\right)^{1/4}\left[\sin(\sqrt{15}\cdot\tau+\omega)\right]^{-1/5}\times$$$$
\times\left[\tan(\sqrt{15}\cdot\tau/2+\omega/2)\right]^{\pm\frac{1}{5}\cdot\sqrt{1+\frac{k^2}{c_1^2+c_2^2}}-\frac{k}{\sqrt{5}\sqrt{c_1^2+c_2^2}}},$$
\begin{equation}a_3=a_0^{3/2}(c_1^2+c_2^2)^{1/4}\left(\frac{q^2_m}{6\mu h_\eta^2}\right)^{-1/2}\left[\sin(\sqrt{15}\cdot\tau+\omega)\right]^{3/5}\times$$$$\times\left[\tan(\sqrt{15}\cdot\tau/2+\omega/2)\right]^{\pm\frac{2}{5}\cdot\sqrt{1+\frac{k^2}{c_1^2+c_2^2}}}.\end{equation}
For the "+"\, sign in (\ref{lofact}), the Universe volume  increases with time ($\alpha'_\tau>0$). The moment $\sqrt{15}\cdot\tau=-\omega$ corresponds to a cosmological initial singularity:
$$a\propto (\sqrt{15}\cdot\tau+\omega)^{\frac{1}{15}\left(1+4\sqrt{1+\frac{k^2}{c_1^2+c_2^2}}\right)} \rightarrow 0,$$
\begin{equation}\label{aom}\dot{\alpha}\propto 1/ (\sqrt{15}\cdot\tau+\omega)\rightarrow\infty.\end{equation}
The energy density of the magnetic field also has the singularity:
\begin{equation}\label{aom1}s_1\propto (\sqrt{15}\cdot\tau+\omega)^{\frac{4}{5}\left(1-\sqrt{1+\frac{k^2}{c_1^2+c_2^2}}\right)} \rightarrow \infty.\end{equation}
For LRS Bianchi-I space-time ($k=0$)  the energy density does not have the singularity at the beginning: $s_1\approx$ const.

In finite time $\sqrt{15}\cdot\Delta\tau=\pi$ the model reaches the big rip singularity:
$$a\propto1/(\pi-\omega-\sqrt{15}\cdot\tau)^{\frac{1}{15}\left(-1+4\sqrt{1+\frac{k^2}{c_1^2+c_2^2}}\right)}\rightarrow\infty,$$
$$\dot{\alpha}\propto 1/(\pi-\omega-\sqrt{15}\cdot\tau)\rightarrow\infty \,\,
\text{as} \,\, \sqrt{15}\cdot\tau\rightarrow\pi-\omega -0.$$
For values $\sqrt{15}\cdot\tau\rightarrow\pi-\omega -0$ the energy density
decreases according to the law
$$s_1\propto (\pi-\omega-\sqrt{15}\cdot\tau)^{\frac{4}{5}\left(1+\sqrt{1+\frac{k^2}{c_1^2+c_2^2}}\right)}\rightarrow 0.$$
 Since the following holds: $\dot{\beta_-}/\dot{\alpha}=(\beta_-)'_\tau/\alpha'_\tau=k/(e^{3\alpha}s_1\alpha'_\tau)\rightarrow \text{const}$  as $\sqrt{15}\cdot\tau\rightarrow\pi-\omega -0$, the anisotropy level $\sigma_\alpha$ does not tend to zero with the expansion of the Universe (see (\ref{levelsd4})).  There is no isotropization process.

For the "$-$"\, sign in (\ref{lofact}), the Universe evolution begins with infinite value $a(\tau)$ and it contracts in a finite time $\Delta\tau=\pi/\sqrt{15}$. The energy density of the magnetic field increases from zero to a non-infinite value at the moment of the Universe collapse.

Although the model describes the Universe with a short life, it has some value. The exact solution provides information about the initial singularity.
The type of singularity may be similar for the model (\ref{mmsd1}), (\ref{mmsd2}), (\ref{mmsd3}) with $l\neq 0$ in case of condition $\varepsilon_l+s\approx s$ as $\tau \rightarrow \tau_{in}$.

\section{Case $l\neq 0$}

Here we note some general properties of the model with $l\neq 0$. From equation (\ref{mmsd3}) it follows that $\alpha'_\tau\cdot\dfrac{s'_\tau}{s}<0$. Therefore, at the moments of the Universe expansion ($\alpha'_\tau>0$), the energy density of the magnetic field decreases ($s'_\tau<0$). Condition (\ref{bnaal}) on the anisotropy level $\sigma_\alpha$ (see (\ref{levelsd4})) gives restrictions
\begin{equation}\left|\frac{s'_\tau}{4s\alpha'_\tau}+1\right|\leq |A|,\,\, \frac{|k|}{|l|\cdot e^{3\alpha}|\varepsilon_l+s||\alpha'_\tau|}\leq |B|,\,\, A^2+B^2<1,\end{equation}
where $A$, $B$ -- constants. This means that $\dfrac{s'_\tau}{s\alpha'_\tau}$ and $ (e^{3\alpha}(\varepsilon_l+s)\alpha'_\tau)^{-1}$ are bounded functions. Next we will consider the issues of singularity, isotropization and inflation. We will look for models of only an expanding Universe.

\subsection{Singularities}

Let us examine the initial singularity and the behavior of the model around it. There are various options depending on the parameters.

I. Knowing the exact solution of the model and its singularities with $l=0$, we assume the existence of a singularity at the bounded point $\tau_0$ of the form
\begin{equation}\label{sing1}\alpha'_\tau\approx\dfrac{\alpha_0}{\tau-\tau_0},\,\, a\approx a_0(\tau-\tau_0)^{\alpha_0}\,\, \text{as} \,\, \tau\rightarrow\tau_0+,\end{equation} where inequalities $\tau\geq\tau_0$, $\alpha_0>0$ provide the expansion of the Universe ($\alpha'_\tau>0$). Since $\dfrac{s'_\tau}{s\alpha'_\tau}$ is the bounded function, it is natural to assume that
\begin{equation}\frac{s'_\tau}{s}\approx -m^2(\tau-\tau_0)^{n},\, \,n\geq -1,\, \,m^2\neq0,\end{equation}
where the choice of the factor ($-m^2$) follows from condition $\alpha'_\tau\cdot\dfrac{s'_\tau}{s}<0$.

1. In case $n=-1$ we have
\begin{equation}\label{sing2}s\approx s_0(\tau-\tau_0)^{-m^2}.\end{equation}
Since $(e^{3\alpha}(\varepsilon_l+s)\alpha'_\tau)^{-1}$ is the bounded function and
$(e^{3\alpha}(\varepsilon_l+s)\alpha'_\tau)^{-1}\propto (\tau-\tau_0)^{-(3\alpha_0-1-m^2)}$ as $\tau\rightarrow\tau_0+$, then $3\alpha_0-1-m^2\leq0$. Under assumptions (\ref{sing1}) and (\ref{sing2}), the function $(e^{3\alpha}(\varepsilon_l+s)\alpha'_\tau)^{-1}$ will always be finite on interval $\tau\geq\tau_0$ in case $\varepsilon_l=1$ ($l>0$).
Assumptions (\ref{sing1}) and (\ref{sing2}) are checked by substituting into the equations (\ref{mmsd2}), (\ref{mmsd3}).
We leave the leading terms on the right and left sides of equation (\ref{mmsd2}):
\begin{equation}m^2(3\alpha_0-1-m^2)(\tau-\tau_0)^{3\alpha_0-m^2-2}\simeq 12(\tau-\tau_0)^{3\alpha_0-m^2},\end{equation}
where the left side dominates at $\tau\rightarrow\tau_0+$. Therefore,
\begin{equation}\label{mkv} m^2=3\alpha_0-1\Rightarrow \alpha_0>1/3.\end{equation}
 In equation (\ref{mmsd3}), terms with $(\tau-\tau_0)^{-2}$ dominate:
\begin{equation}\left(-(3\alpha_0-1)^2+8\alpha_0(3\alpha_0-1)-\frac{16k^2}{|l|^2a_0^6s_0^2}\right)(\tau-\tau_0)^{-2}\simeq0,\end{equation}
then
\begin{equation}15\alpha_0^2-2\alpha_0-1-\frac{16k^2}{|l|^2a_0^6s_0^2}=0.\end{equation}
One of the roots of this equation is positive:
\begin{equation}\alpha_0=\frac{1}{15}+\frac{4}{15}\cdot\sqrt{1+\frac{15k^2}{|l|^2a_0^6s_0^2}}.\end{equation}
Condition $\alpha_0>1/3$ is satisfied only for $k\neq0$.
From equality (\ref{mkv}) follows $m^2=\frac{4}{5}\left(-1+\sqrt{1+\frac{15k^2}{|l|^2a_0^6s_0^2}}\right)$.
Thus, we obtained approximation:
$$a\propto (\tau-\tau_0)^{\frac{1}{15}+\frac{4}{15}\cdot\sqrt{1+\frac{15k^2}{|l|^2a_0^6s_0^2}}}\rightarrow 0,$$
\begin{equation}s\propto (\tau-\tau_0)^{\frac{4}{5}\left(1-\sqrt{1+\frac{15k^2}{|l|^2a_0^6s_0^2}}\right)}\rightarrow \infty \,\, \text{as}\,\, \tau\rightarrow\tau_0+.\end{equation}
The behavior near the singularity is similar to (\ref{aom}), (\ref{aom1}) (the model with $l=0$).

Using Eq. (\ref{mmsd1}), (\ref{ajn}), we can determine the anisotropy factors $e^{\beta_{\pm}}$:
\begin{equation}(\beta_-)'_\tau\approx\frac{k}{|l|a_0^3s_0}\cdot\frac{1}{(\tau-\tau_0)}\Rightarrow e^{\beta_-}\approx b_-(\tau-\tau_0)^{\frac{k}{|l|a_0^3s_0}},\end{equation}
$$(\beta_+)'_\tau\approx-\frac{\alpha_0+1}{4(\tau-\tau_0)}=-\frac{1}{15(\tau-\tau_0)}\left(4+\sqrt{1+\frac{15k^2}{|l|^2a_0^6s_0^2}}\right)\Rightarrow$$
\begin{equation}\Rightarrow e^{\beta_+}\approx b_+(\tau-\tau_0)^{-\frac{1}{15}\left(4+\sqrt{1+\frac{15k^2}{|l|^2a_0^6s_0^2}}\right)}.\end{equation}
Then we can write the three scale factors
$$a_{1,2}=e^{\alpha+\beta_{+}\pm \sqrt{3}\beta_-}\approx a_0b_{+}b_{-}^{\pm\sqrt{3}}(\tau-\tau_0)^{\frac{1}{5}\left(-1+\sqrt{1+\frac{15k^2}{|l|^2a_0^6s_0^2}}\right)\pm\frac{\sqrt{3}k}{|l|a_0^3s_0}},$$
\begin{equation}\label{a1a2a3d} a_{3}=e^{\alpha-2\beta_+}\approx a_0b_{+}^{-2}(\tau-\tau_0)^{\frac{1}{5}\left(3+2\sqrt{1+\frac{15k^2}{|l|^2a_0^6s_0^2}}\right)}.\end{equation}
The constants of integration are related by the equality $a_0b_{+}\cdot s_0^{1/4}=\left(\frac{q^2_m}{6|l|\mu h_\eta^2}\right)^{1/4}.$
For all values of $k$, the scale factor $a_3$ has a limit $a_3\rightarrow 0$ as $\tau\rightarrow\tau_0+$. For $k>0$ we have limits: $a_1\rightarrow 0$, $a_2\rightarrow +\infty$;
for $k<0$: $a_1\rightarrow +\infty$, $a_2\rightarrow 0$. Thus there is a thread-like singularity at the beginning $\tau_0$.

 The obtained approximation is valid for $k\neq 0$ ($a_1\neq a_2 \neq a_3$).

2. In case $n>-1$ we have
\begin{equation}s\approx s_0\left[1-\frac{m^2}{n+1}\cdot(\tau-\tau_0)^{n+1}\right].\end{equation}
For $\varepsilon_l+s_0=0$ this asymptotic approximation is not valid. Next we assume $\varepsilon_l+s_0\neq 0$. Due to the arbitrariness of $s_0>0$,
it follows that $\varepsilon_l=1$ ($l>0$). From limitation $(e^{3\alpha}(\varepsilon_l+s)\alpha'_\tau)^{-1}$ it follows that $\alpha_0\leq 1/3$.
Let's substitute approximate functions into the equations (42), (\ref{mmsd3}):
$$m^2(3\alpha_0+n)(\tau-\tau_0)^{n-1+3\alpha_0}\simeq 12(\tau-\tau_0)^{3\alpha_0},$$
\begin{equation}8\alpha_0m^2(\tau-\tau_0)^{n-1}\simeq 16+m^4(\tau-\tau_0)^{2n}+\frac{16k^2}{|l|^2(1+s_0)^2}\cdot(\tau-\tau_0)^{-6\alpha_0}.\end{equation}
The assumed approximation is valid for $k=0$ (LRS model), $n=1$, $m^2=6$, $\alpha_0=1/3$:
\begin{equation}s\approx s_0[1-3(\tau-\tau_0)^2]\rightarrow s_0,\,\, a\approx a_0(\tau-\tau_0)^{1/3}\rightarrow 0\,\, \text{as}\,\, \tau\rightarrow\tau_0+.\end{equation}
For case $k=0$, in our work \cite{Muhryamov4q} an exact expression was obtained
$$-4h_\eta\cdot (\pm t)=\sqrt{\frac{5}{3}}\arcsin\left[\frac{1}{4}\cdot\sqrt{\frac{3}{5}}(5 + s)\right]+\ln s-$$\begin{equation}\label{tst}-
\ln\left[5-15 s+\sqrt{5}\sqrt{5-30  s-3s^2}\right]+c.\end{equation}

In this case there is a pancake singularity:
\begin{equation}a_{1}=a_{2}\approx\left(\frac{q^2_m}{6\mu h_\eta^2|l|s_0}\right)^{1/4}, \,\,
a_3\approx \left(\frac{6\mu h_\eta^2|l|s_0}{ q^2_m}\right)^{1/2} \cdot b_0\cdot (h_\eta\cdot t)\rightarrow 0\,\, \text{as}\,\, \tau\rightarrow\tau_0+,\end{equation}
where $s_0=4\cdot \sqrt{\frac{5}{3}}-5\approx 0.164$.

II. The model allows an approximation at the bounded point $\tau_0$:
\begin{equation}\alpha'_\tau\approx \alpha_0(\tau-\tau_0)^{n}, \,\, \frac{s'_\tau}{s}\approx -m^2(\tau-\tau_0)^{n},\,\, \text{as} \,\, \tau\rightarrow\tau_0+.\end{equation}
Therefore
\begin{equation}a\approx a_0\left[1+\frac{\alpha_0}{n+1}\cdot(\tau-\tau_0)^{n+1}\right] ,\,\,s\approx s_0\left[1-\frac{m^2}{n+1}\cdot(\tau-\tau_0)^{n+1}\right].\end{equation}
Inequalities $\alpha_0>0$, $-1<n<0$ provide the expansion of the Universe.
This approximation is valid for $\varepsilon_l+s_0=0$, therefore $\varepsilon_l=-1$ ($l<0$), $s_0=1$.
We leave the leading terms on the right and left sides of equation (\ref{mmsd2}):
\begin{equation}m^2(2n+1)(\tau-\tau_0)^{2n}\simeq12(\tau-\tau_0)^{n+1},\end{equation}
where the left side dominates at $\tau\rightarrow\tau_0+$. Therefore,
\begin{equation}\label{mkv1}n=-1/2.\end{equation}
In equation (\ref{mmsd3}), terms with $(\tau-\tau_0)^{-1}$ dominate:
\begin{equation}\left(-8\alpha_0m^2+m^4+\frac{4k^2}{|l|^2a_0^6m^4}\right)(\tau-\tau_0)^{-1}\simeq0,\end{equation}
then
\begin{equation}\alpha_0=\frac{m^2}{8}+\frac{k^2}{2|l|^2a_0^6m^6}.\end{equation}
Thus, we obtained approximation:
$$a\approx a_0\left[1+\left(\frac{m^2}{4}+\frac{k^2}{|l|^2a_0^6m^6}\right)(\tau-\tau_0)^{1/2}\right],$$
\begin{equation}s\approx 1-2m^2(\tau-\tau_0)^{1/2}.\end{equation}
The Universe evolution begins with a non-zero value.

From (\ref{mmsd1}) it follows
\begin{equation}(\beta_-)'_\tau\approx-\frac{k}{|l|a_0^3m^2}\cdot\frac{1}{2(\tau-\tau_0)^{1/2}}\Rightarrow\beta_-\approx-\frac{k}{|l|a_0^3m^2}\cdot(\tau-\tau_0)^{1/2}+\text{const}.\end{equation}
From (\ref{ajn}) it follows
$$(\beta_+)'_\tau\approx\frac{-\alpha_0+m^2/4}{(\tau-\tau_0)^{1/2}}=\left(\frac{m^2}{4}-\frac{k^2}{|l|^2a_0^6m^6}\right)\cdot\frac{1}{2(\tau-\tau_0)^{1/2}}\Rightarrow$$
\begin{equation}\Rightarrow\beta_+\approx\left(\frac{m^2}{4}-\frac{k^2}{|l|^2a_0^6m^6}\right)\cdot(\tau-\tau_0)^{1/2}+\text{const}.\end{equation}
The scalar factors $a_i$ start dynamics from a non-zero value:
$$a_{1,2}\approx a_0b_{+}b_{-}^{\pm\sqrt{3}}\left(1+c_{1,2}\cdot(\tau-\tau_0)^{1/2}\right),$$
\begin{equation}a_{3}\approx  a_0b_{+}^{-2}\left(1+c_{3}\cdot(\tau-\tau_0)^{1/2}\right),\,\, c_i=\text{const}.\end{equation}

Three combinations of parameters ($k\neq0$, $l>0$); ($k=0$, $l>0$); ($l<0$) determine different behavior of the model near the starting point $\tau_0$. Thus, for $l>0$, LRS model and the general model ($a_1\neq a_2 \neq a_3$) have different approximations at $\tau\rightarrow\tau_0$.

\subsection{The isotropization process}

It was previously noted that isotropization ($\sigma_\alpha\rightarrow 0$) will occur at $|\dot{\alpha}|\rightarrow h_\eta$ or $|\alpha'_\tau|\rightarrow 1$. If the Universe is expanding, there will be
 $\alpha'_\tau\rightarrow 1$. Therefore (see (\ref{levelsd4})), $a(\tau)\approx a_\infty e^{\tau}$ and
\begin{equation}\frac{s'_\tau}{4s\alpha'_\tau}+1\rightarrow 0,\,\, \frac{k^2}{|l|^2\cdot e^{6\alpha}\left(\varepsilon_l+s\right)^2}
\cdot\frac{1}{\alpha'^2_\tau}\rightarrow 0.\end{equation}
Then the approximation follows
\begin{equation}\frac{s'_\tau}{s}\approx -4 \Rightarrow s\approx s_\infty e^{-4\tau},\,\, \tau\gg 1.\end{equation}
Next, we will check whether the system of equations allows this approximation.
We will add terms of a higher order of smallness to the approximation:
$$\frac{s'_\tau}{s}\approx -4+Ae^{-4\tau}+Ce^{-6\tau},$$
\begin{equation}\alpha'_\tau\approx 1+Be^{-4\tau}+De^{-6\tau}.\end{equation}
Let's substitute the equation into (\ref{mmsd3}):
$$Be^{-4\tau}+De^{-6\tau}-\frac{k^2}{2|l|^2a_\infty^6}\cdot e^{-6\tau}=0\Rightarrow$$
\begin{equation} B=0,\,\, D=\frac{k^2}{2|l|^2a_\infty^6}.\end{equation}
From equation (\ref{mmsd2}), we have
\begin{equation}A=16\varepsilon_ls_\infty,\,\,C=-\frac{2k^2}{|l|^2a_\infty^6}.\end{equation}
Thus
$$\frac{s'_\tau}{s}\approx -4+16\varepsilon_ls_\infty\cdot e^{-4\tau} -\frac{2k^2}{|l|^2a_\infty^6}\cdot e^{-6\tau}\rightarrow -4,$$
\begin{equation}\alpha'_\tau\approx 1+\frac{k^2}{2|l|^2a_\infty^6}\cdot e^{-6\tau}\rightarrow 1\,\, \text{as}\,\, \tau\rightarrow +\infty.\end{equation}
The assumed approximation is valid.
From (\ref{mmsd1}), (\ref{ajn}) it follows
$$(\beta_-)'_\tau\approx\frac{k}{|l|a_\infty^3e^{3\tau}}\Rightarrow \beta_-\approx\text{const}_{-}- \frac{k}{3|l|a_\infty^3e^{3\tau}},$$
\begin{equation}(\beta_+)'_\tau\approx -4\varepsilon_ls_\infty e^{-4\tau}\Rightarrow \beta_+\approx\text{const}_{+}+\varepsilon_ls_\infty e^{-4\tau}.\end{equation}
Taking into account (\ref{levelsd4}) we obtain an approximation $|\sigma_\alpha|\propto e^{-3\tau}\rightarrow 0$ as $\tau\rightarrow +\infty$ ($k\neq0$).
 For comparison, the LRS model ($k=0$) has an approximation $|\sigma_\alpha|\propto e^{-4\tau}\rightarrow 0$, i.e.  the anisotropy decreases faster.
For the scalar factors the following holds: $a_{i}(\tau)\approx a_\infty^{(i)}e^\tau$. The limit $a_i/a\rightarrow \text{const}_i$ confirms the isotropization process of the model. The isotropization process is allowed for different signs of $\varepsilon_l$ ($l\lessgtr 0$). The space-time becomes isotropic with time and it is de Sitter world ($\alpha'_\tau\rightarrow 1$).

\subsection{Selecting parameter values for the model}

Here we present reasons for choosing the parameter values in the model.

The model becomes isotropic over time and is approximately determined by the equalities:
\begin{equation}\label{etta23}a(\tau)\approx a_\infty e^{\tau}=a_\infty e^{h_\eta\cdot t},\,\,
\dot{\phi}^2=2X\approx 3\varepsilon\mu(\varepsilon_\Lambda h_\Lambda^2-h_\eta^2),\,\, H=h_\eta.\end{equation}
In the isotropic cosmological model two conditions related
to scalar perturbations must be satisfied \cite{DeFelice1}:
\begin{equation}c_S^2\equiv\frac{3(2w_1^2w_2H-w^2_2w_4+4w_1w_2\dot w_1-2w_1^2\dot w_2)}{w_1(4w_1w_3+9w_2^2)}\geq 0\,,\label{cS}
\end{equation}
for the avoidance of Laplacian instabilities associated with the
scalar field propagation speed, and
\begin{equation}
Q_{S}\equiv\frac{w_{1}(4w_{1}w_{3}+9w_{2}^{2})}{3w_{2}^{2}}>0\,,
\label{QS}
\end{equation}
for the absence of ghosts. For the tensor perturbations the conditions for avoidance of ghost
and Laplacian instabilities are respectively written as
\cite{DeFelice1}
\begin{equation}Q_T\equiv\frac{w_1}{4}>0\,,\, c^2_{T}\equiv\frac{w_4}{w_1}\geq 0\,.\label{csT}
\end{equation}
{ \it Where $w_i$ have the form:}
\begin{eqnarray}
&&w_{1}  \equiv  2\,(G_{{4}}-2\,
XG_{{4X}})-2X\,(G_{{5X}}{\dot{\phi}}H-G_{{5\phi}}) \,,
\label{w1def}\\
&&w_{2}  \equiv  -2\, G_{{3X}}X\dot{\phi}+4\,
G_{{4}}H-16\,{X}^{2}G_{{4{ XX}}}H+4(\dot{\phi}G_{ {4\phi X}}-4H\,
G_{{4X}})X+2\, G_{{4\phi}}\dot{\phi} \nonumber
\\
 &  & \ \ \ \ \ \ \ \,
 +8\,{X}^{2}HG_{{5\phi X}}+2H\, X\,(6G_{{5\phi}}-5\,
G_{{5X}}\dot{\phi}{H})-4G_{{5{
XX}}}{\dot{\phi}}X^{2}{H}^{2}\,,\\
&& w_{3}  \equiv  3\, X(G_{2{X}}+2\, XG_{{
2XX}})+6X(3X\dot{\phi}HG_{{3{ XX}}}-G_{{3\phi X}}
X-G_{{3\phi}}+6\, H\dot{\phi}G_{{3X}})
\nonumber \\
 &  & \ \ \ \ \ \ \ \,
 +18\, H(4\, H{X}^{3}G_{{4{  XXX}}}-HG_{{4}}-5\, X\dot{\phi}G_{{4\phi
X}}-G_{{4\phi}}\dot{ \phi}+7\, HG_{{4X}}X+16\, H{X}^{2}G_{{4{
XX}}}-2\,{X}^{2}\dot{\phi}G_{{4\phi{
XX}}})\nonumber \\
 &  & \ \ \ \ \ \ \ \,    +6{H}^{2}X(2\, H\dot{\phi}G_{{5{
XXX}}}{X}^{2}-6\,{X}^{2}G_{{5\phi{  XX}}}+13XH\dot{\phi}G_{{5{
XX}}}-27G_{{5\phi
X}}X+15\, H\dot{\phi}G_{{5X}}-18G_{{5\phi}})\,,\\
&&w_{4}  \equiv  2G_{4}-2XG_{5\phi}-2XG_{5X}\ddot{\phi}~.
\end{eqnarray}

For the subclass of Horndeski theory (\ref{lagr1}), (\ref{g2}) functions $w_i$ have the form:
$$w_1=\mu+\eta X,\, w_2=2H(\mu+3\eta X),$$\begin{equation}w_3=3(\varepsilon X-3\mu H^2-18\eta XH^2),\, w_4=\mu-\eta X.\end{equation}
{\it Here, we took into account: $G_2=\varepsilon X-\Lambda$, $G_3=0$, $G_4=\mu/2$, $G_5=\eta\cdot\phi/2$. }
In case of the approximation (\ref{etta23}) and the condition (\ref{ettavar}) they are transformed as follows
$$w_1=\frac{\mu}{2}\left[1+\varepsilon_\Lambda\frac{h^2_\Lambda}{h^2_\eta}\right],\, w_2=\mu h_\eta\left[-1+3\varepsilon_\Lambda\frac{h^2_\Lambda}{h^2_\eta}\right],$$
\begin{equation}\label{w1234}w_3=\frac{9}{2}\mu h^2_\eta\left[3-5\varepsilon_\Lambda\frac{h^2_\Lambda}{h^2_\eta}\right],\, w_4=\frac{\mu}{2}\left[3-\varepsilon_\Lambda\frac{h^2_\Lambda}{h^2_\eta}\right].\end{equation}
Then the conditions $c^2_S\geq0$, $Q_S>0$, $c^2_T\geq0$, $Q_T>0$ will give inequalities
$$c^2_S=\frac{1}{3}\cdot \frac{-1+3\varepsilon_\Lambda\cdot\frac{h^2_\Lambda}{h^2_\eta}}{1+\varepsilon_\Lambda\cdot\frac{h^2_\Lambda}{h^2_\eta}}\geq0,$$
$$Q_S=6\mu\left[1+\varepsilon_\Lambda\cdot\frac{h^2_\Lambda}{h^2_\eta}\right]\cdot \left[\frac{1-\varepsilon_\Lambda\cdot\frac{h^2_\Lambda}{h^2_\eta}}{-1+3\varepsilon_\Lambda\cdot\frac{h^2_\Lambda}{h^2_\eta}}\right]^2>0,$$
\begin{equation}\label{cqts}c^2_T=\frac{3-\varepsilon_\Lambda\cdot\frac{h^2_\Lambda}{h^2_\eta}}{1+\varepsilon_\Lambda\cdot\frac{h^2_\Lambda}{h^2_\eta}}\geq0,\,\,
Q_T=\frac{\mu}{8}\left[1+\varepsilon_\Lambda\cdot\frac{h^2_\Lambda}{h^2_\eta}\right]>0.\end{equation}
From $Q_S>0$, $Q_T>0$ it follows $l=1+\varepsilon_\Lambda\cdot\frac{h^2_\Lambda}{h^2_\eta}>0$. Then $c^2_S\geq0$ and $c^2_T\geq0$ lead to
\begin{equation}\label{sdf3} \Lambda>0 \, (\varepsilon_\Lambda>0),\, 1/3<\frac{h^2_\Lambda}{h^2_\eta}\leq 3.\end{equation}
The presence of the parameter $\Lambda$ is a necessary condition for the stability of the model. Indeed, in case  $\Lambda=0$ there is Laplace instability: $c^2_S=-1/3<0$.
Observed data  on the tensor-to-scalar ratio \begin{equation}r=\mathcal{P}_T/\mathcal{P}_S=4Q_S/Q_T\end{equation} allow us to refine estimates (\ref{sdf3}).  Here $\mathcal{P}_S$ -- the power spectrum of the scalar perturbation, $\mathcal{P}_T$ -- the power spectrum of the tensor perturbations. Сonstraints from the PLANCK observations \cite{Aghanim} at the moment estimated as
\begin{equation}\label{tentoscalar} r=192\left[\frac{1-\frac{h^2_\Lambda}{h^2_\eta}}{-1+3\frac{h^2_\Lambda}{h^2_\eta}}\right]^2<r_0=0.1. \end{equation}
Therefore,
\begin{equation}\label{tentoscalar11} \left[\frac{1+(r_0/192)^{1/2}}{1+3(r_0/192)^{1/2}}\right]^{1/2}<\frac{h_\Lambda}{h_\eta}<
\left[\frac{1-(r_0/192)^{1/2}}{1-3(r_0/192)^{1/2}}\right]^{1/2},\,\,\text{or}\,\, 0.978<\frac{h_\Lambda}{h_\eta}<1.024.\end{equation}
  In the case of constraints from Planck/WMAP/BICEP/Keck \cite{BICEP} ($r<r_0=0.036$), we have the estimate:
\begin{equation}0.987<\frac{h_\Lambda}{h_\eta}<1.014.\end{equation}

The $\Lambda$-term provides a parametric degree of freedom that allows the model to be tuned to the observed data.

\section{Case $l>0$, $\Lambda>0$}
In the previous section we presented arguments in favor of choice $l> 0$, $\Lambda>0$. Here we will present the solution of field equations with parameters $l> 0$, $\Lambda>0$ on the interval $\tau \in (\tau_0,+\infty)$. Numerical methods were used in the solution process.

\subsection{Model without magnetic field}
First, we consider the influence of NMKC without the magnetic field on the space-time in case $l>0$, $\Lambda>0$, $\varepsilon>0$.
We put $q_m=0$, and the consequence of the field equations becomes
\begin{equation}\label{pcj}\dot{\phi}^2=3\mu(h_\Lambda^2-h_\eta^2),\end{equation}
$$\dot{\beta}_{\pm}=\frac{k_{\pm}}{e^{3\alpha}},\,\, k_{\pm}=\text{const},$$
\begin{equation}\label{bezmagn}\dot{\alpha}^2=h_\eta^2+\dot{\beta_+}^2+\dot{\beta_-}^2=h_\eta^2+\frac{\sigma_0^2}{e^{6\alpha}},
\end{equation}
where $\sigma_0^2\equiv k^2_++k^2_-$.
The model (\ref{bezmagn}) is similar to the well-known with $L=\sqrt{-g}\left(\mu R/2-\mathbf{\Lambda}^*\right)$ in the Bianchi I metric, where $\mathbf{\Lambda}^*=\varepsilon\mu/\eta>0$.
The anisotropy term $\sigma_0^2/e^{6\alpha}$ in the last equation decreases with time, hence anisotropy contribution becomes irrelevant and the Universe rapidly approaches the isotropic de Sitter phase with parameter $H=h_\eta$. The anisotropy terms become dominant at small $e^{\alpha}$, when one can neglect the $h^2_\eta$-term and the Universe is described by the Kasner metric:
$a_i\propto t^{p_i}$, $p_1+p_2+p_3=p^2_1+p^2_2+p^2_3=1$. The model has an initial singularity at $t\rightarrow 0$: $a=e^\alpha\propto (h_\eta\cdot t)^{1/3} \rightarrow 0$, $\dot{\alpha}\propto 1/t\rightarrow\infty.$ The anisotropic behavior is standard: the anisotropy is significant near the initial singularity and decreases monotonically with time.

Metric functions of the model (\ref{bezmagn}) do not contain the $\Lambda$-term; it is replaced by $\eta$: $|\Lambda|/\mu\rightarrow \varepsilon/\eta$. In other words, the $\Lambda$-term is screened. The driving force of primary inflation is  NMKC through parameter $\eta$. Only the scalar field (\ref{pcj}) "feels"\, the $\Lambda$-term.

\subsection{Model with the magnetic field}

\begin{figure}[h]
\includegraphics[width=10cm]{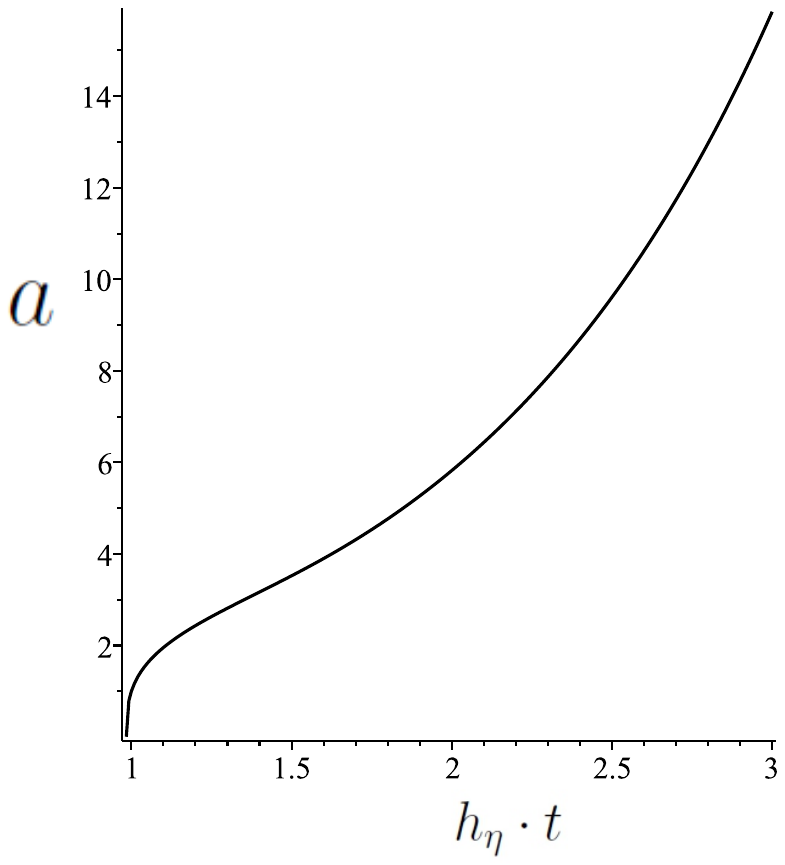}
\caption{The profile  $a(\tau)$. The starting point is $\tau_0\approx 0.984315$.  The scalar factor  $a(\tau)=e^{\alpha(\tau)}$ is an increasing function, i.e. the Universe volume increases with time.  \label{a}}
\end{figure}

\begin{figure}[h]
\includegraphics[width=10cm]{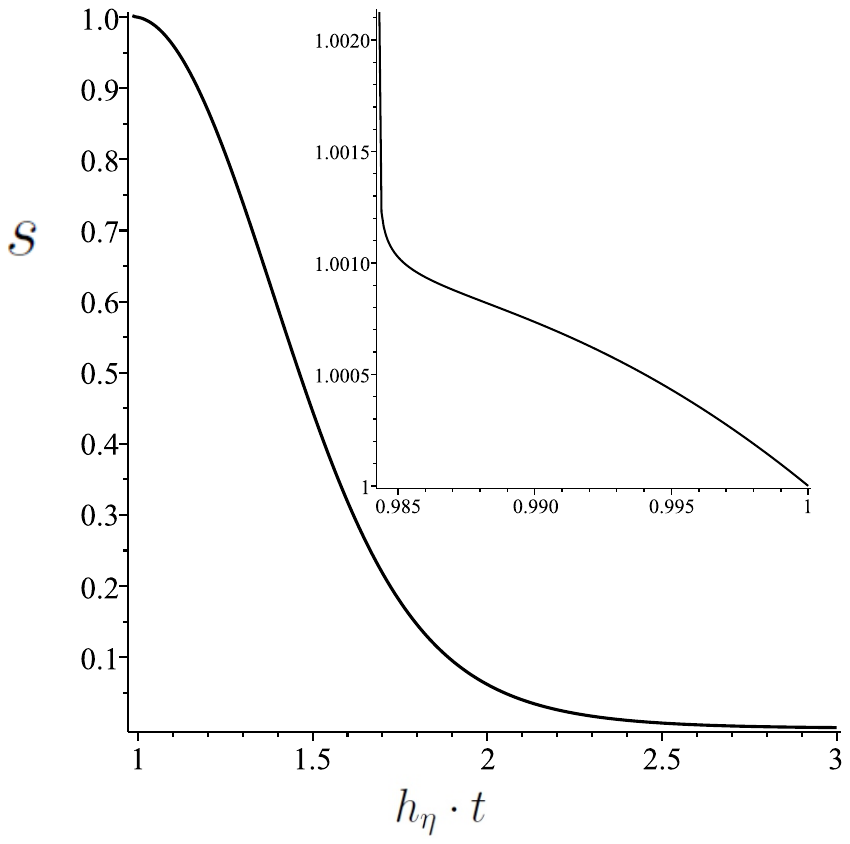}
\caption{The profile  $s(\tau)$. The dimensionless function $s(\tau)$ is associated with the magnetic field energy density $\mathcal{E}^{({\rm em})}$: $s=\frac{1}{3\mu h_\eta^2|l|}\cdot\mathcal{E}^{({\rm em})}$. The graph shows that the magnetic energy density is a monotonic decreasing function. \label{b}}
\end{figure}

\begin{figure}[h]
\includegraphics[width=10cm]{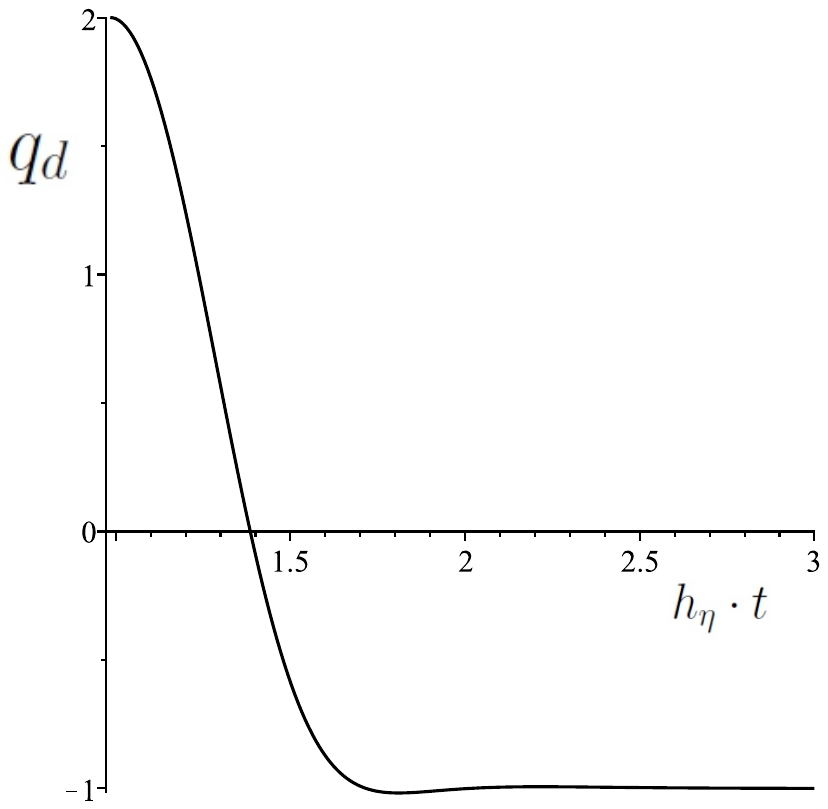}
\caption{The profile  $q_d(\tau)$. The starting point is $\tau_0\approx 0.984315$. There are two phases. In the first phase,
there is no acceleration ($q_d\geq0$), and the second one is characterized by the acceleration expansion of the Universe ($q_d<0$). \label{c}}
\end{figure}

\begin{figure}[h]
\includegraphics[width=10cm]{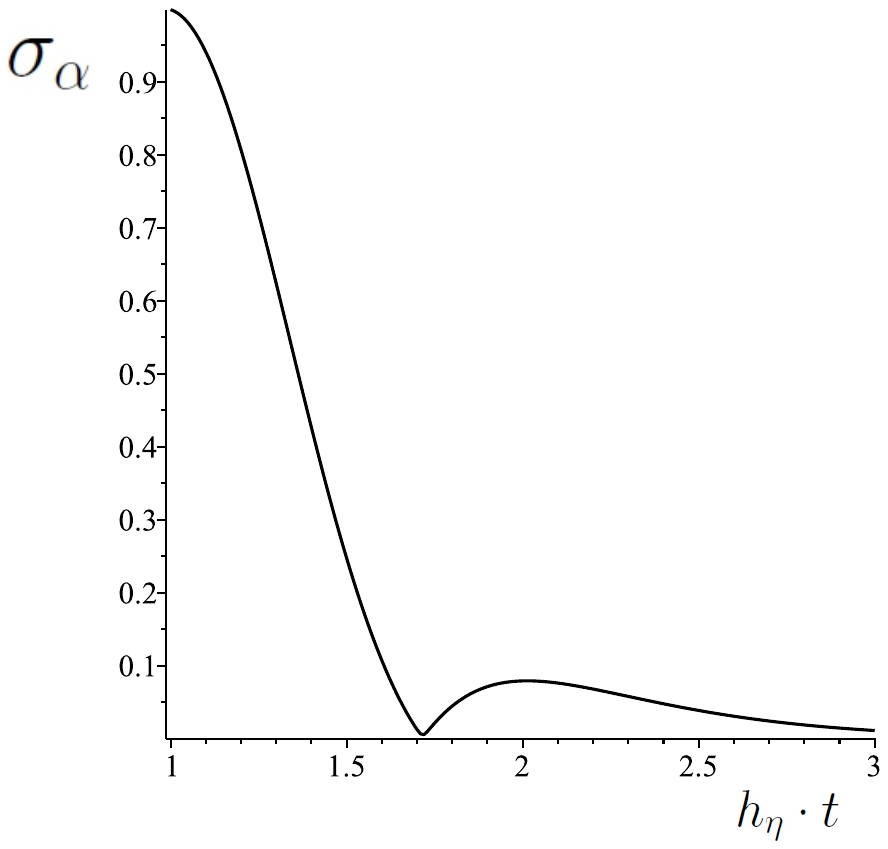}
\caption{The profile  $\sigma_\alpha(\tau)$. There is a rebound point on the plot. The anisotropy level is a non-monotonic function. This is the non-standard  behavior.  But, in the end, the process of isotropization occurs: $\sigma_\alpha\propto e^{-3\tau}\rightarrow 0$, $\tau\rightarrow +\infty$. Within a few units of $\tau=h_\eta \cdot t$, the anisotropy becomes small, therefore $a(\tau)\propto e^{\tau}$ will be almost throughout the entire inflation time interval. \label{d}}
\end{figure}

\begin{figure}[h]
\includegraphics[width=10cm]{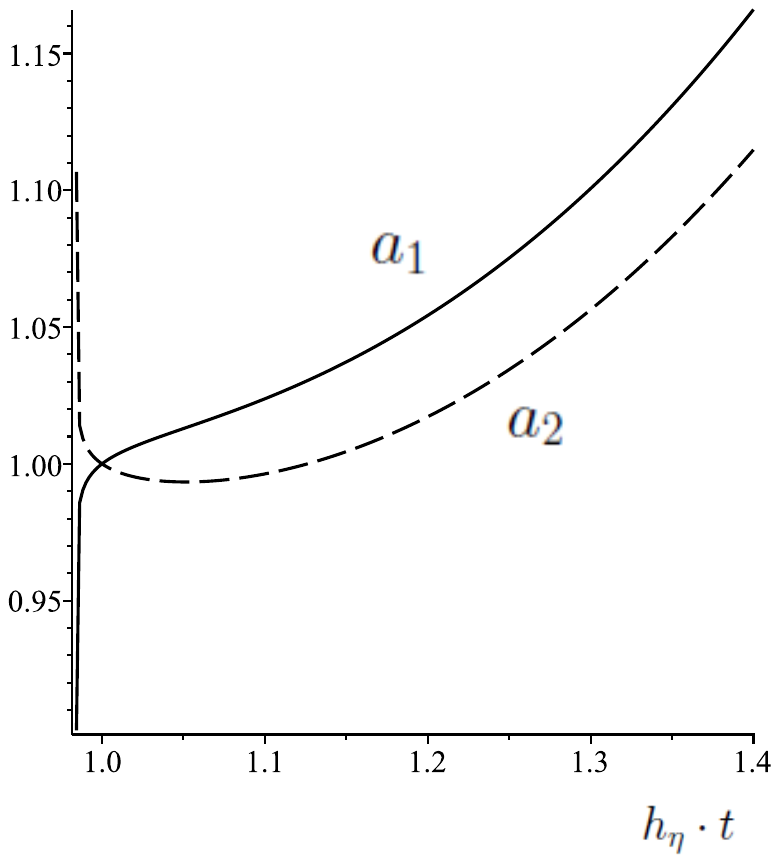}
\caption{The profile  $a_1(\tau)$, $a_2(\tau)$ for $k>0$. The starting point is $\tau_0\approx 0.984315$.  The figure shows a bouncing scale factor $a_2$.  At some point in time, the Universe begins to expand in all directions. \label{e1}}
\end{figure}

\begin{figure}[h]
\includegraphics[width=10cm]{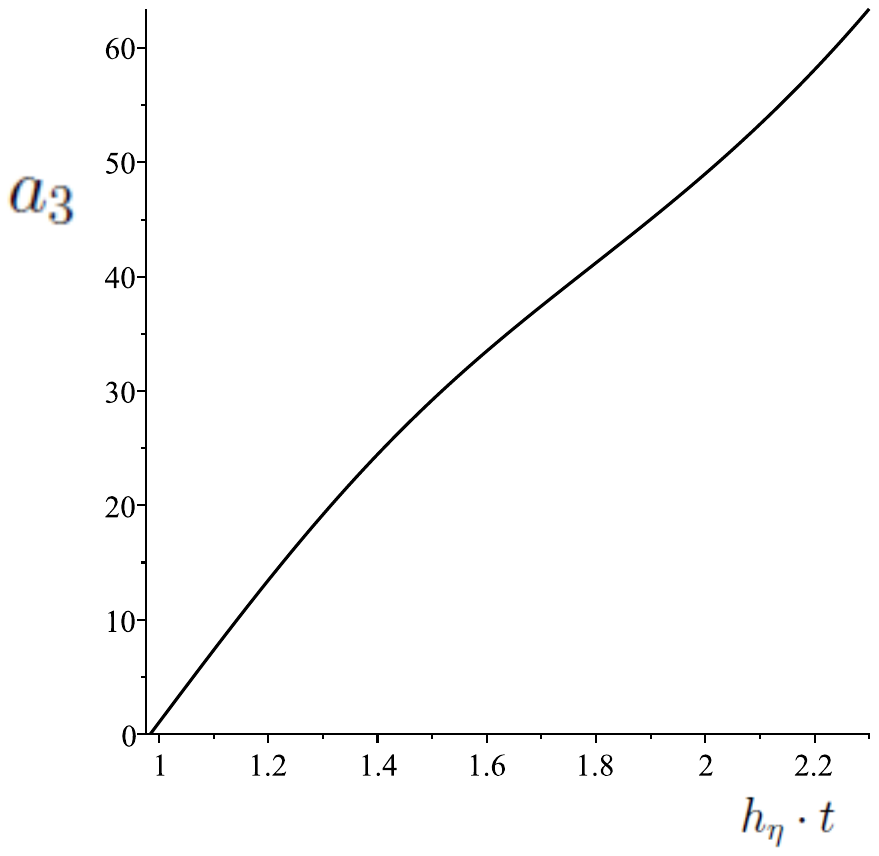}
\caption{The profile  $a_3(\tau)$ for $k>0$. The starting point is $\tau_0\approx 0.984315$. The scale factor increases monotonically over time. \label{e2}}
\end{figure}

Here, we will consider the combined effect of the magnetic field and NMKC on the space-time.
Using (\ref{mmsd2})-(\ref{mmsd3}), we can plot the graph Fig.\ref{a} for the scale factor $a(\tau)=e^{\alpha(\tau)}$ and the graph Fig.\ref{b} for the magnetic field energy density $s(\tau)$, when $k\neq0$. The starting point is $\tau_0\approx 0.984315$.
The Universe volume  increases with time. From figure Fig.\ref{c} it is clear that the deceleration parameter (DP) evolves from positive values at past epoch to negative
values at late time:
\begin{equation}q_d=\frac{d}{dt}\left(\frac{1}{\dot{\alpha}}\right)-1.\end{equation}
The Universe begins to expand without acceleration. Later comes a phase of accelerated expansion. According to the model, the primary inflation occurs at time $\tau-\tau_0\approx 0.4$, and it should last at least 60 Hubble times (e-folds), $\triangle\tau=h_\eta\triangle t\approx 60$. That is the post-singularity era is much shorter than the inflation period.

The magnetic energy density is a  monotonic decreasing function. In case $k\neq0$, $l>0$ function $s(\tau)$ changes in interval $(0,+\infty)$. This result allows us to obtain constraints on the model parameters from condition $\dot{\phi}^2>0$. From (\ref{phemlp}) (exact expression) it follows
\begin{equation}\label{phemlp12}(\phi'_\tau)^2=3\varepsilon\mu\left[-1+ \frac{h^2_\Lambda}{h^2_\eta}+\left(1+\frac{h^2_\Lambda}{h^2_\eta}\right)\cdot s\right].\end{equation}
In limit $s\rightarrow +\infty$, the equality $\dot{\phi}^2>0$ will be satisfied if $\varepsilon=1>0$. The phantom scalar field is not allowed in the presented model. In limit  $s\rightarrow 0$ we obtain inequality $\frac{h_\Lambda}{h_\eta}>1$.
Inequalities (\ref{tentoscalar11}) are clarified:
\begin{equation}\label{nj45}1<\frac{h_\Lambda}{h_\eta}<\left[\frac{1-(r_0/192)^{1/2}}{1-3(r_0/192)^{1/2}}\right]^{1/2}\approx1.024,\,\, \text{where}\,\, r_0=0.1.\end{equation}
In case $k=0$, $l>0$ we get the same estimate (\ref{nj45}). The phantom scalar field ($\varepsilon=-1$) will contradict the estimate (\ref{tentoscalar11}) (see \cite{Muhryamov4q}).
Thus, for all $k$ we get
$$\Lambda>0,\,\, \eta>0,\,\, \varepsilon=1,$$
\begin{equation}\label{njk2} 1<\Lambda\eta/\mu<\frac{1-(r_0/192)^{1/2}}{1-3(r_0/192)^{1/2}}\approx 1.049,\,\, \text{where}\,\, r_0=0.1. \end{equation}
 For $r_0=0.036$:
\begin{equation}\label{njk21} 1<\Lambda\eta/\mu<\frac{1-(r_0/192)^{1/2}}{1-3(r_0/192)^{1/2}}\approx 1.028. \end{equation}
NMKC and $\Lambda$-term are similar on the energy scale: $\mu h_\eta^2\approx\mu h_\Lambda^2$, i.e $l\approx 2$.
Under conditions (\ref{njk2}), (\ref{njk21}), the model will describe the Universe that enters the inflationary regime without the pathologies of instability.

The behavior of the anisotropy level $\sigma_\alpha$ can be seen in Fig.\ref{d}. There is a rebound point on the plot. The anisotropy level is a non-monotonic function. This is the non-standard  behavior. But, in the end, the process of isotropization occurs: $\sigma_\alpha\propto e^{-3\tau}\rightarrow 0$, $\tau\gg 1$. Moreover, the model becomes almost isotropic in a short period of time ($\triangle\tau\approx 5$)
compared to the duration of the primary inflation ($\triangle\tau\approx 60$). Therefore $a(\tau)\approx a_\infty e^\tau$ will be almost throughout the entire inflation time interval. For model (\ref{bezmagn}) without the magnetic field and LRS model (see \cite{Muhryamov4q}), the anisotropy level is а monotonically decreasing function.

Scale factors $a_i(t)$ are shown in Fig. \ref{e1}, \ref{e2} for $k>0$. The figure shows a bouncing scale factor $a_2$. The function $a_2$ falls from a greater value at the beginning, bounce the minimum value, and then rise again at the end. Other scale factors $a_{1,3}$ start dynamics from zero. At some point in time, the Universe begins to expand in all directions. In case $k<0$, the scale factors $a_{1,2}$ change places ($a_1\leftrightarrow a_2$). From (\ref{a1a2a3d}), we see that near the initial singularity, the Universe is not described by the Kazner metric. For the model without the magnetic field and LRS model (see \cite{Muhryamov4q}), there is the Kazner epoch.

Thus, in case $k\neq 0$ (i.e $a_1\neq a_2\neq a_3$), parameters $l>0$, $\Lambda>0$, $\eta>0$, $\varepsilon=1$ provide the desired properties of the model: early isotropization, rapid transition from the post-singularity era to inflation, absence of ghost and Laplace instabilities. It is possible to "harmoniously"\, include the magnetic field in NMKC.

\section{Conclusion}

In this paper, we obtain cosmological solutions within the framework of Bianchi I model with a unidirectional magnetic field in the theory of gravity with non-minimum kinetic coupling of a scalar field with curvature (\ref{lagr1}). We believe that the solutions are applicable to the period before and during primary inflation. In the previous work \cite{Muhryamov4q} we considered such a problem for a special case -- the locally rotationally symmetric (LRS) Bianchi-I space-time ($a_1=a_2$). Here we generalize the results for different scale factors $a_1\neq a_2 \neq a_3$. There are distinctive properties of the models and a number of interesting properties are preserved.

{\bf I.} These models have the following common properties.
\begin{enumerate}
  \item The zero scalar charge $C_\phi=0$ leads to a limited level of anisotropy. Its dynamics develops in a limited region  $\sigma_\alpha<1$.
A necessary condition for isotropization of models is the same signs of parameters  $\varepsilon$ and $\eta$.

  \item If the magnetic field is not included, then NMKC reduces to the model with $L=\sqrt{-g}\left(\mu R/2-\mathbf{\Lambda}^*\right)$ in the Bianchi I metric, where $\mathbf{\Lambda}^*=\varepsilon\mu/\eta>0$. Anisotropy suppresses NMKC and the Universe is described by the Kasner metric. After a short period of time, the influence of NMKC becomes significant and the Universe rapidly approaches the isotropic de Sitter phase with parameter $h_\eta=\sqrt{\frac{\varepsilon}{3\eta}}$. The initial $\Lambda$-term is screened: it is replaced by $\eta$: $|\Lambda|/\mu\rightarrow \varepsilon/\eta$. Only the scalar field "feels"\, the $\Lambda$-term.

  \item Due to the inclusion of the magnetic field in NMKC, the sign of the parameter $l=1+\varepsilon\eta \Lambda/\mu$  begins to qualitatively influence the properties of the model.
  \begin{itemize}
    \item In case $l=0$, the Universe has a short life. Either the Universe collapses in a finite time or reaches the big rip singularity. There is no isotropization process.

    \item In case $l>0$, the model can describe the expansion of the Universe over an infinite time interval. The magnetic energy density decreases over time. After a short post-singularity epoch, within a few $h_\eta\cdot t$ units, the anisotropy becomes small. The process of isotropization occurs and the Universe enters the quasi-de Sitter epoch with the parameter $h_\eta$. The model with the magnetic field has the same asymptotic behavior (\ref{etta23}) at $h_\eta\cdot t=\tau\gg 1$ as NMKC without the magnetic field. Constraints $\Lambda>0$, $\eta>0$, $\varepsilon=1$, $1<\Lambda\eta/\mu<\frac{1-(r_0/192)^{1/2}}{1-3(r_0/192)^{1/2}}\approx 1.049$ ensure the observed tensor-scalar ratio $r<r_0=0.1$  (for $r_0=0.036$: $1<\Lambda\eta/\mu<1.028$), and the absence of ghosts, Laplace instability. The phantom scalar field is not allowed in the presented model.

    \item In case $l<0$,  the model has ghost and Laplacian instabilities.
  \end{itemize}

\end{enumerate}

{\bf II.} The LRS model ($k=0$) and the model $a_1\neq a_2 \neq a_3$ ($k\neq 0$) have the following differences.

\begin{enumerate}
  \item The magnetic energy density:
  \begin{itemize}
  \item in case  $k=0$, for all $l$ the magnetic energy density decreases with the expansion of the Universe from a bounded value;

  \item in case  $k\neq 0$, for values $l\geq0$ the magnetic energy density decreases from a infinite value;

  \item for $l<0$, the magnetic energy density takes on limited values.

\end{itemize}

  \item  The singularity.

  \begin{itemize}
  \item Set ($k=0$, $l>0$). There is the pancake singularity at the beginning; the Universe is described by the Kasner metric ($p_1=p_2=0$, $p_3=1$) in the neighborhood of the initial point.
  \item Set ($k\neq 0$, $l>0$). There is the thread-like singularity at the beginning; there is no the Kasner metric.
  \item Set ($l<0$) -- the scalar factors $a_i$ start dynamics from a non-zero value.
\end{itemize}

  \item The anisotropy level.

  For ($k=0$, $l>0$), the anisotropy level is а monotonically decreasing function ($\sigma_\alpha\propto e^{-4\tau}\rightarrow 0$, $\tau\gg 1$), and for ($k\neq 0$, $l>0$) it is non-monotonic. In the latter case there is a rebound point on the plot. This is the non-standard  behavior.
But, in the end, the process of isotropization occurs: $\sigma_\alpha\propto e^{-3\tau}\rightarrow 0$, $\tau\gg 1$.  For ($k=0$, $l>0$)  the anisotropy decreases faster.

\end{enumerate}

Thus, for model $k\neq 0$, the main conclusion remains valid. For $l>0$, the model has the necessary properties: isotropization during expansion, rapid transition to inflationary expansion, absence of ghost and Laplace instabilities, i.e. the magnetic field does not cause pathologies. In other cases $l\leqslant 0$, the model has various disadvantages.

The models under consideration do not contain a mechanism for exiting primordial inflation. We artificially limit the model by specifying its period of applicability. In our case, this is approximately $h_\eta\cdot t=60$.  Therefore, our next step will be to model the transition from primordial inflation to the next stage of the Universe's evolution. To this end, we will include ordinary matter in the model and possibly complicate the electromagnetic component by considering the Born-Infeld theory.

\begin{acknowledgements}
This work was funded by a grant from the Academy of Sciences of the Republic of Tatarstan provided to higher education institutions, scientific and other organizations to support human resource development plans in terms of encouraging their research and academic staff to defend doctoral dissertations and conduct research activities (Agreement No. 12/2025-PD-KFU dated December 22, 2025).
\end{acknowledgements}

\end{document}